\documentstyle[12pt,epsf]{article}

\catcode`@=11
\def\chkspace{%
  \relax   
  \begingroup\ifhmode\aftergroup\dochksp@ce\fi\endgroup}
\def\dochksp@ce{%
  \unskip              
  \futurelet\chkspct@k\d@chkspc  
}
\def\d@chkspc{%
  \let\nxtsp@ce=\relax
  \ifx\chkspct@k.\else     
    \ifx\chkspct@k,\else
      \ifx\chkspct@k;\else
        \ifx\chkspct@k!\else
          \ifx\chkspct@k?\else
            \ifx\chkspct@k:\else
              \ifx\chkspct@k)\else
              \ifx\chkspct@k(\else
                \ifx\chkspct@k]\else
                  \ifx\chkspct@k-\else
                    \ifx\chkspct@k\egroup\else  
                      \let\nxtsp@ce=\put@space  
                    \fi
                  \fi
                \fi
              \fi
              \fi
            \fi
          \fi
        \fi
      \fi
    \fi
  \fi
  \nxtsp@ce
}
\def\put@space{$\;$}
\catcode`@=12

\def\ra{{$\rightarrow$}\chkspace}

\def\etal{{\it et al.}\chkspace}

\def\ep{{e$^+$e$^-$}\chkspace}

\def\gluino{\relax\ifmmode \tilde{g} \else $\tilde{g}$ \fi\chkspace}

\chkspace
\chkspace
\def\m0{$M_{0}$}\chkspace
\def\m0m{$M_{0}MAX$}\chkspace
\chkspace
\chkspace

\def\bbrm{\relax\ifmmode {\rm b}\bar{\rm b}
       \else ${\rm b}\bar{\rm b}$ \fi\chkspace}
\def\bb{$b\bar{b}$ \chkspace}
\def\ccrm{\relax\ifmmode {\rm c}\bar{\rm c}
       \else ${\rm c}\bar{\rm c}$ \fi\chkspace}
\def\cc{$c\bar{c}$ \chkspace}
\def\tt{\relax\ifmmode {\rm t}\bar{\rm t}
       \else ${\rm t}\bar{\rm t}$ \fi\chkspace}
\def\ss{\relax\ifmmode {\rm s}\bar{\rm s}
       \else ${\rm s}\bar{\rm s}$ \fi\chkspace}
\def\uu{\relax\ifmmode {\rm u}\bar{\rm u}
       \else ${\rm u}\bar{\rm u}$ \fi\chkspace}
\def\dd{\relax\ifmmode {\rm d}\bar{\rm d}
       \else ${\rm d}\bar{\rm d}$ \fi\chkspace}

\def\qqg{\relax\ifmmode {\rm q}\overline{\rm q}{\rm g}
\else q$\overline{\rm q}$g \fi\chkspace}

\def\afb{\relax\ifmmode A_{FB} \else
{{$A_{FB}$}}\fi\chkspace}
\def\afbb{\relax\ifmmode A_{FB}^b \else
{{$A_{FB}^b$}}\fi\chkspace}
\def\pafb{\relax\ifmmode \tilde{A}_{FB} \else
{{$\tilde{A}_{FB}$}}\fi\chkspace}
\def\pafbb{\relax\ifmmode \tilde{A}_{FB}^b \else
{{$\tilde{A}_{FB}^b$}}\fi\chkspace}

\def\pafbzo{\relax\ifmmode \tilde{A}_{FB}|_{O(0)} \else
{{$\tilde{A}_{FB}|_{O(0)}$}}\fi\chkspace}
\def\pafbfo{\relax\ifmmode \tilde{A}_{FB}|_{\oalp} \else
{{$\tilde{A}_{FB}|_{\oalp}$}}\fi\chkspace}
\def\pafbso{\relax\ifmmode \tilde{A}_{FB}|_{\oalpsq} \else
{{$\tilde{A}_{FB}|_{\oalpsq}$}}\fi\chkspace}
\def\pafbto{\relax\ifmmode \tilde{A}_{FB}|_{\oalpc} \else
{{$\tilde{A}_{FB}|_{\oalpc}$}}\fi\chkspace}

\def\pafbbzo{\relax\ifmmode \tilde{A}_{FB}^b|_{O(0)} \else
{{$\tilde{A}_{FB}^b|_{O(0)}$}}\fi\chkspace}
\def\pafbbfo{\relax\ifmmode \tilde{A}_{FB}^b|_{\oalp} \else
{{$\tilde{A}_{FB}^b|_{\oalp}$}}\fi\chkspace}
\def\pafbbso{\relax\ifmmode \tilde{A}_{FB}^b|_{\oalpsq} \else
{{$\tilde{A}_{FB}^b|_{\oalpsq}$}}\fi\chkspace}
\def\pafbbto{\relax\ifmmode \tilde{A}_{FB}^b|_{\oalpc} \else
{{$\tilde{A}_{FB}^b|_{\oalpc}$}}\fi\chkspace}

\def\afbo0{\tilde{A}_{FB}|_{O(0)}}
\def\afbo1{\tilde{A}_{FB}|_{\oalp}}
\def\afbo2{\tilde{A}_{FB}|_{\oalpsq}}
\def\afbo3{\tilde{A}_{FB}|_{\oalpc}}

\def\lam{\relax\ifmmode \Lambda_{\overline{MS}}
       \else {{$\Lambda_{\overline{MS}}$}}\fi\chkspace}
\def\lamuds{\relax\ifmmode \Lambda^{(3)}_{\overline{MS}}
       \else {{$\Lambda^{(3)}_{\overline{MS}}$}}\fi\chkspace}
\def\lamudsc{\relax\ifmmode \Lambda^{(4)}_{\overline{MS}}
       \else $\Lambda^{(4)}_{\overline{MS}}$\fi\chkspace}
\def\lamudscb{\relax\ifmmode \Lambda^{(5)}_{\overline{MS}}
       \else $\Lambda^{(5)}_{\overline{MS}}$\fi\chkspace}

\def\alp{\relax\ifmmode \alpha_s\else $\alpha_s$\fi\chkspace}
\def\alpbar{\relax\ifmmode \bar{\alpha_s}
       \else $\bar{\alpha_s}$\fi\chkspace}
\def\alpmz{\relax\ifmmode \alpha_s(M_Z)\else $\alpha_s(M_Z)$\fi\chkspace}
\def\alpmzsq{\relax\ifmmode \alpha_s(M_Z^2)
       \else $\alpha_s(M_Z^2)$\fi\chkspace}

\def\oalp{\relax\ifmmode O(\alpha_s)\else{{O($\alpha_s$)}}\fi\chkspace}
\def\oalpsq{\relax\ifmmode O(\alpha_s^2)
           \else{{O($\alpha_s^2$)}}\fi\chkspace}
\def\oalpc{\relax\ifmmode O(\alpha_s^3)
           \else{{O($\alpha_s^3$)}}\fi\chkspace}
\def\oalpf{\relax\ifmmode O(\alpha_s^4)
           \else{{O($\alpha_s^4$)}}\fi\chkspace}

\def\rb{\relax\ifmmode R_3^b/R_3^{all}
           \else{{$R_3^b/R_3^{all}$}}\fi\chkspace}
\def\rc{\relax\ifmmode R_3^c/R_3^{all}
           \else{{$R_3^c/R_3^{all}$}}\fi\chkspace}
\def\ruds{\relax\ifmmode R_3^{uds}/R_3^{all}
           \else{{$R_3^{uds}/R_3^{all}$}}\fi\chkspace}
\def\ri{\relax\ifmmode R_3^i/R_3^{all}
           \else{{$R_3^i/R_3^{all}$}}\fi\chkspace}
\def\rj{\relax\ifmmode R_3^j/R_3^{all}
           \else{{$R_3^j/R_3^{all}$}}\fi\chkspace}
\def\alpi{\relax\ifmmode \alpha^i_s/\alpha^{all}_s
           \else{{$\alpha^i_s/\alpha^{all}_s$}}\fi\chkspace}

\def\prl{Phys. Rev. Lett.\chkspace}
\def\prd{Phys. Rev.\chkspace}

\def\z0{{$Z^0$}\chkspace}
\def\Dst{\relax\ifmmode {\rm D}^* \else {D$^*$}\fi\chkspace}
\def\Dpl{\relax\ifmmode {\rm D}^+ \else {D$^+$}\fi\chkspace}
\def\D0{\relax\ifmmode {\rm D}^0 \else {D$^0$}\fi\chkspace}
\def\Kst{\relax\ifmmode {\rm K}^* \else {K$^*$}\fi\chkspace}
\def\K0{\relax\ifmmode {\rm K}^0_s \else {K$^0_s$}\fi\chkspace}
\def\Kpl{\relax\ifmmode {\rm K}^+ \else {K$^+$}\fi\chkspace}
\def\Kstz{\relax\ifmmode {\rm K}^{*0} \else {K$^{*0}$}\fi\chkspace}

\def\ep{{$e^+e^-$}\chkspace}

\def\z0{$Z^0$}

\def\bb{{$b\bar{b}$}\chkspace}
\def\cc{{$c\bar{c}$}\chkspace}

\def\etal{{\it et al.}\chkspace}

\def\qqg{{$q\bar{q}g$}\chkspace}

\renewcommand{\baselinestretch}{1.0}

 1

\catcode`\@=11 
\makeatletter
\def\@seccntformat#1{\csname the#1\endcsname.\hskip 1em}
\makeatother

\begin{document}

\thispagestyle{empty}
\begin{flushright}
{\renewcommand{\baselinestretch}{.75}
  SLAC--PUB--9966\\
September 2003\\
}
\end{flushright}

\vskip 0.5truecm
 
\begin{center}
{\large\bf First Measurement of the Double-inclusive $B/\overline{B}$ 
Hadron Energy Distribution in \ep Annihilations, and of 
Angle-Dependent Moments of the $B$ and $\overline{B}$ Energies$^*$
}
\end{center}
\vspace {0.4cm}

\begin{center}
 {\bf The SLD Collaboration$^{**}$}\\
Stanford Linear Accelerator Center \\
Stanford University, Stanford, CA~94309
\end{center}
 
\vspace{0.5cm}
 
\begin{center}
{\bf ABSTRACT }
\end{center}

\noindent
We have made the first measurement 
of the double-inclusive $B$/$\overline{B}$ energy distribution in \ep
annihilations, using a sample of 400,000 hadronic \z0-decay events recorded 
in the SLD experiment at SLAC between 1996 and 1998.  
The small and stable SLC beam spot 
and the CCD-based vertex detector were used to reconstruct 
$B$/$\overline{B}$-decay vertices with high efficiency and purity, 
and to provide precise measurements of the kinematic
quantities used to calculate the $B$ energies in this novel technique.  
We measured the $B$/$\overline{B}$ 
energies
with good efficiency and resolution over the full kinematic range. 
We measured moments of the scaled energies of the
$B$ and $\overline{B}$ hadrons vs. the opening angle between them. 
By comparing these results with
perturbative QCD predictions we tested the ansatz of factorisation in 
heavy-quark production. A recent next-to-leading order calculation reproduces 
the data. 

\vspace {1.5cm}
 
\vfill
{\footnotesize
$^*$ Work supported in part by Department of Energy contract DE-AC03-76SF00515.}

\eject

\rm  
\section{Introduction}
\noindent 
The production of heavy hadrons ($H$) in \ep annihilation provides a
laboratory for the study of heavy-quark ($Q$) jet fragmentation. This is 
commonly characterized in terms of the observable 
$x_{H}$ $\equiv$ $2E_H/\sqrt{s}$, where
$E_H$ is the energy of a $B$ or $D$ hadron containing a $b$ or $c$ quark,
respectively, and $\sqrt{s}$ is the c.m. energy. 
The distribution of $x_H$, $D(x_H)$, is conventionally referred to as the
heavy-quark `fragmentation function'\footnote{Unless stated otherwise, in
these studies we do
not distinguish between hadrons and antihadrons.}.

In recent publications we presented~\cite{sldoldbfrag,sldnewbfrag}
the results of a new method for reconstructing
$B$-hadron decays, and the $B$ energy, inclusively, using only charged tracks,
in the SLD experiment at SLAC.
We used the upgraded charge-coupled device (CCD) vertex detector, 
installed in 1996, to reconstruct $B$-decay vertices in \z0 decays with high 
efficiency and purity.  Combined with the micron-sized SLC interaction point
(IP), our precise vertexing allowed us to reconstruct the total
transverse momentum of the tracks from $B$-decays, and therefore the transverse
momentum and mass associated with the neutral particles in the $B$-decays.
This allowed us to reconstruct accurately
the energy of $B$ hadrons. These studies 
yielded the  most precise measurement of the $b$-quark fragmentation 
function, and allowed us to test models of heavy-quark fragmentation. Of the
9 models tested, only 4 were consistent with our precision data at better than 
the 1\% level based on a $\chi^2$ probability. This allowed us to reduce the
model-dependent systematic uncertainty on the $b$-quark fragmentation function.

We have extended these studies and applied similar `topological' vertexing 
techniques to tag events in which  we reconstructed the energies of
both leading $B$ hadrons produced via \ep \ra \bb \ra $B\overline{B} + X$.
We measured the moments of the single-inclusive $B$-hadron scaled-energy
distribution $dN/dx_{B}$:
\begin{equation}
D_i\quad\equiv\quad \int x_B^{i-1} \frac{1}{N_s}\;\frac{dN_s}{dx_B} \;d x_B
\label{eqn:single}
\end{equation}
as well as the moments of the double-inclusive scaled-energy distribution:
\begin{equation}
D_{ij}(\phi)\quad\equiv\quad \int\int x_{B1}^{i-1} x_{B2}^{j-1} \frac{1}{N_d}\;
\frac{d^3N_d}
{dx_{B1} dx_{B2} d\cos\phi} \;dx_{B1} dx_{B2}, 
\label{eqn:double}
\end{equation}
where $x_{B1}$ and $x_{B2}$ are the scaled energies of the two $B$ hadrons and
the label is arbitrary, $\phi$ is the angle between their flight 
directions, and $i$ and $j$ are integers $\geq$1.
We formed the normalised moments:
\begin{equation}
G_{ij}(\phi)\quad\equiv\quad D_{ij}(\phi)/(D_i D_j)
\label{eqn:norm}
\end{equation}
Following the method proposed in~\cite{phil} we used 
these quantities to test the ansatz of 
factorisation as
applied to perturbative  Quantum Chromodynamics (pQCD) calculations of 
\ep \ra \bb events.

Due to `soft', or 
non-perturbative, effects, $D_i$ and $D_{ij}$ cannot be predicted 
absolutely in pQCD.
Rather, the respective pQCD calculation must be folded with models of the 
non-perturbative (`hadronisation') process in order to derive predictions
that can be compared with experimental data.
 However, 
provided that the ansatz of factorisation holds~\cite{phil}, 
namely that calculation of the perturbative and non-perturbative phases can be
separated by (an arbitrary) factorisation scale $\mu_F$, the dependence on
$\mu_F$ cancels in Eq.~\ref{eqn:norm} and hence $G_{ij}$ can be
calculated absolutely in pQCD, up to possible `higher twist' effects of
order $1/\sqrt{s}$, with no dependence on hadronisation models. 
Comparison of the measured
$G_{ij}$ with pQCD predictions hence allows both a test of this ansatz and of 
the perturbative calculations. 
The $G_{ij}$ can be derived at leading order (LO) in pQCD using the numerical
results in~\cite{phil}. In addition, next-to-leading order (NLO) predictions 
for $G_{ij}$ have been calculated recently~\cite{bno}. 

In Section 2 we describe the detector and the selection of \ep \ra hadrons 
events used in this analysis. We present in Section 3 the first measurement of
the double-inclusive $B$-hadron energy distribution and, in Section 4, 
 of the normalised 
moments Eq.~\ref{eqn:norm}. In Section 5 we describe the estimation
of the errors on our measurements. Finally, in Section 6, we test the
ansatz of factorisation and
compare our data with the pQCD predictions.
 
\section{Apparatus and Hadronic Event Selection}
 
\noindent
This analysis is based on roughly 400,000 hadronic \z0 events produced in 
e$^+$e$^-$ annihilations at a mean center-of-mass energy of 
$\sqrt{s}=91.28$ GeV
at the SLAC Linear Collider (SLC), and recorded in the SLC Large Detector
(SLD) between 1996 and 1998. 
A general description of the SLD can be found elsewhere~\cite{sld}.
The trigger and initial selection criteria for hadronic $Z^0$ decays are 
described in Ref.~\cite{sldalphas}.
This analysis used charged tracks measured in the Central Drift
Chamber (CDC)~\cite{cdc} and in the upgraded Vertex Detector (VXD3)~\cite{vxd}.
Momentum measurement was enabled by a uniform axial magnetic field of 0.6T.
The CDC and VXD3  gave a momentum resolution of
$\sigma_{p_{\perp}}/p_{\perp}$ = $0.01 \oplus 0.0026p_{\perp}$,
where $p_{\perp}$ is the track momentum transverse to the beam axis in
GeV/$c$. In the plane normal to the beamline 
the centroid of the micron-sized SLC IP was reconstructed from tracks
in sets of approximately thirty sequential hadronic \z0 decays to a precision 
of $\sigma_{IP}^{r\phi}\simeq4$ $\mu$m.  The IP position along the 
beam axis was determined event by event, using charged tracks, with 
a resolution of $\sigma_{IP}^z$ $\simeq$ 20 $\mu$m.
Including the uncertainty on the IP position, the resolution on the 
charged-track impact parameter ($d$) projected in the plane perpendicular
to the beamline was 
 $\sigma_{d}^{r\phi}$ = 8$\oplus$33/$(p\sin^{3/2}\theta)$ $\mu$m,
and the resolution in the plane containing the beam axis was  
 $\sigma_{d}^{z}$ = 10$\oplus$33/$(p\sin^{3/2}\theta)$ $\mu$m,
where
$\theta$ is the track polar angle with respect to the beamline.

A set of cuts was applied to the data to select well-measured tracks
and events well contained within the detector acceptance.
Charged tracks were required to have a distance of
closest approach transverse to the beam axis within 5 cm,
and within 10 cm along the axis from the measured IP,
as well as $|\cos \theta |< 0.87$, and $p_\perp > 0.15$ GeV/c.
Events were required to have a minimum of five such tracks,
a thrust axis~\cite{thrust} polar angle w.r.t. the beamline, $\theta_T$,
within $|\cos\theta_T|<0.80$, and
a charged visible energy $E_{vis}$ of at least 20~GeV,
which was calculated from the selected tracks assigned the charged pion mass. 
The efficiency for selecting a well-contained $Z^0 \rightarrow q{\bar q}(g)$
event was estimated to be above 97\% independent of quark flavor. The
selected sample comprised 313,447 events, with an estimated
$0.10 \pm 0.05\%$ background contribution dominated
by $Z^0 \rightarrow \tau^+\tau^-$ events.

For the purpose of estimating the efficiency and purity of the
selection procedures we made use of a detailed Monte Carlo (MC) simulation 
of the detector.
The JETSET 7.4~\cite{jetset} event generator was used, with parameter
values tuned to hadronic \ep annihilation data~\cite{tune},
combined with a simulation of $B$-hadron decays
tuned~\cite{sldsim} to $\Upsilon(4S)$ data and a simulation of the SLD
based on GEANT 3.21~\cite{geant}.
Inclusive distributions of single-particle and event-topology observables
in hadronic events were found to be well described by the
simulation~\cite{sldalphas}. Uncertainties in the simulation 
were taken into account in the systematic errors (Section~\ref{sec:sys}). 

\section{$B$-Hadron Selection and Energy Measurement}

The event sample for this analysis was selected using a `topological' vertexing
technique based on the detection and measurement of charged tracks,
which is described in detail in Refs.~\cite{zvnim,dan,gav}. 
We considered events in which we found decay vertices
corresponding to both the leading $B$ and $\overline{B}$ hadrons.
First, the Durham algorithm~\cite{durham} was applied to the selected 
hadronic events, with
a $y_c$ parameter value of 0.015, in order to define a jet structure in 
each event.  
We found that this algorithm and $y_c$ value 
minimized the number of $B$ (and $D$)
decay tracks assigned to the wrong jet.  This is an
important feature for our analysis because we used vertex-related variables 
derived only from charged tracks. Events containing 2, 3, or 4 jets were 
retained for further analysis.

In each selected event, the vertexing algorithm was applied to 
the set of tracks in each jet.
Vertices consistent with photon conversions or $K^{0}$ or $\Lambda^0$ decays 
were discarded.
Events were retained in which a vertex was found in exactly two jets. 35,137
events were selected, of which 89.4\% were estimated to be of \bb origin. 
The efficiency
for selecting true \bb events was estimated to 36.3\%. 

The large masses of the $B$ hadrons relative to light-flavor hadrons 
make it possible to distinguish $B$-hadron decay vertices from those 
vertices found in events of lighter flavors using the vertex invariant 
mass, $M$. However, due to those particles missed from the vertex, 
which are mainly neutrals, $M$ cannot be fully determined.  
$M$ can be written 
\begin{equation}
M=\sqrt{M_{ch}^{2}+P_{t}^{2}+P_{chl}^{2}}+\sqrt{M_{0}^{2}+P_{t}^{2}
\label{eqn:vertexmass}
+P_{0l}^{2}}
\end{equation}
where $M_{ch}$ and $M_{0}$ are the total invariant masses of the set of 
vertex-associated tracks and the set of missing particles, respectively.
$P_{t}$ is the total charged-track momentum transverse to the $B$ flight 
direction, which, by momentum conservation, is identical to the 
transverse momentum of the set of 
missing particles.  $P_{chl}$ and $P_{0l}$ are 
the respective momenta along the $B$ flight direction, which we take to be the
vector joining the IP to the vertex.  

\begin{figure}[ht]      
\vskip 5truecm
\epsfysize5.0 in
\epsfxsize5.5 in
\epsfbox{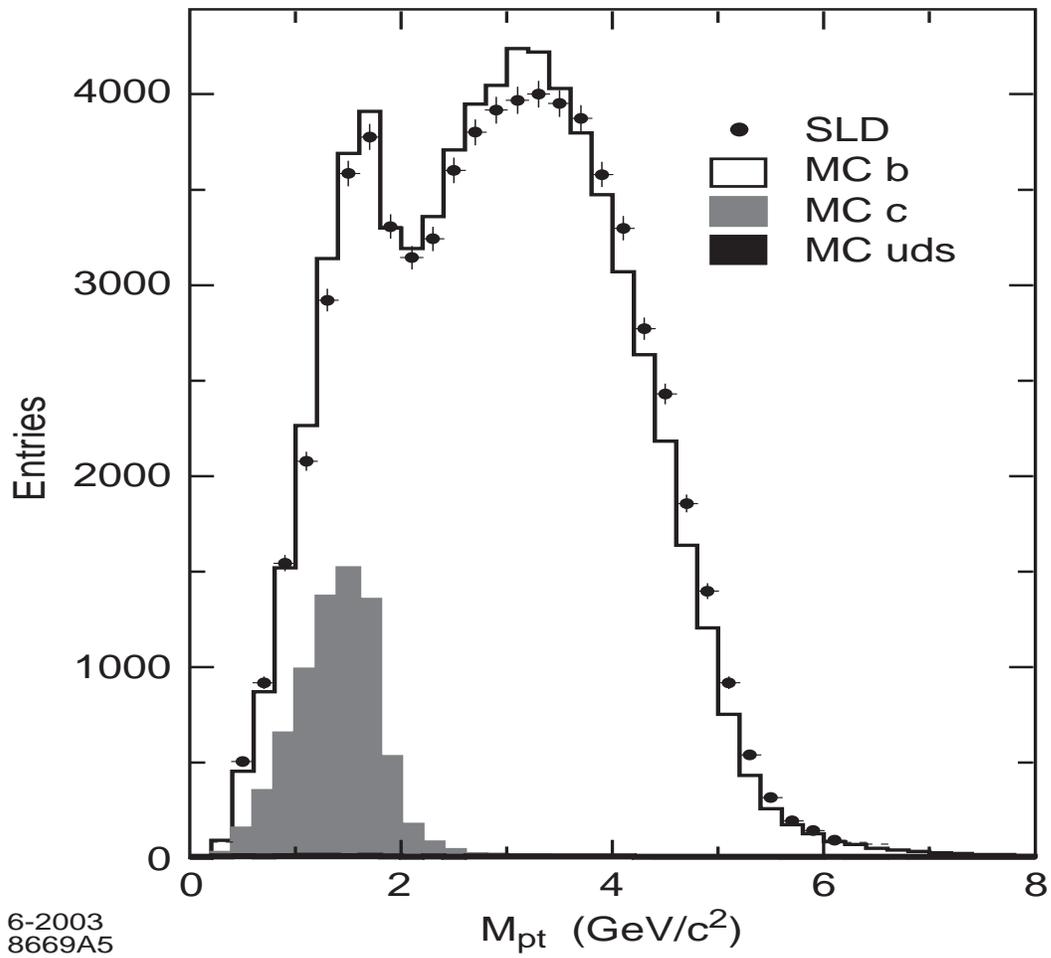}
\caption{$P_t$-corrected mass (see text); data (points) compared with the
simulation (histograms) in which the primary flavor content is indicated.}
\label{fig:pt}
\end{figure}

The lower bound for the mass of the decaying hadron, 
the `$P_{t}$-corrected vertex mass'~\cite{sldnewbfrag}, 
\vspace{-0.2cm}
\begin{equation}
M_{Pt} = \sqrt{M_{ch}^{2}+P_{t}^{2}} + |P_{t}|
\label{eqn:masspt}
\end{equation}
was used as the variable for selecting $B$ hadrons.
Figure~\ref{fig:pt} shows the distribution of $M_{Pt}$ 
for vertices in the selected event sample, and the corresponding simulated 
distribution.
Events were selected that contained at least one vertex with
$M_{Pt}$ $>$ 2.0 GeV/$c^{2}$ and  
$M_{Pt} \leq 2 \times M_{ch}$. The latter cut was found
 to reduce the contamination from fake
vertices in light-quark events.

In order to improve the \bb purity of the sample, 
events were selected in which both vertices had a flight length, $d_{vtx}$, 
such that $0.1 < d_{vtx} < 2.3$ cm; in which
at least one vertex contained two `significant' tracks,
i.e. tracks with a normalised impact-parameter significance, 
$d/\sigma_d$, of at least 2 units; and in which the angle between the vertex
flight vectors, $\phi$, satisfied cos$\phi$ $<$ 0.99. The last cut was
effective at removing 
events in which either a gluon had split into heavy quarks, or the jet-finder 
had artificially split a single heavy-quark jet into two jets.

The energy of each $B$ hadron, $E_{B}$, can be expressed as
the sum of the reconstructed-vertex energy, $E_{ch}$, 
and the energy of those particles not associated with the vertex, $E_{0}$.  
We can write  
\begin{equation}
  E_{0}^{2} =  M_{0}^{2} + P_{t}^{2} + P_{0l}^{2} 
\label{eqn:e0}
\end{equation}
The two unknowns, $M_{0}$ and $P_{0l}$, must be found in order 
to obtain $E_{0}$.
One kinematic constraint can be obtained by imposing the $B$-hadron mass 
on the vertex, $M_{B}^{2}=E_{B}^{2}-P_{B}^{2}$, where 
$P_{B}=P_{chl}+P_{0l}$ is the total momentum of the $B$ hadron. 
From Equation~(\ref{eqn:vertexmass}) we 
derive the following inequality,
\begin{equation}
  \sqrt{M_{ch}^2 + P_{t}^2} + \sqrt{M_{0}^2 + P_{t}^2} \leq M_{B}, 
\label{massineq}
\end{equation}
where equality holds in the limit where
both $P_{0l}$ and $P_{chl}$ vanish in the $B$ hadron rest frame.
Equation~(\ref{massineq}) effectively sets an upper bound on 
$M_{0}$, and a lower bound is given by zero:
\begin{equation}
   0\leq M_{0}^{2}\leq M_{0max}^{2},
\end{equation}
where 
\begin{equation}
M_{0max}^{2}=M_{B}^2 - 2M_{B}\sqrt{M_{ch}^2+P_{t}^2} + M_{ch}^2. 
\label{m0maxeqn}
\end{equation}

Because $M_{0}$ peaks near $M_{0max}$,~\cite{dan} 
we set $M_{0}^{2}$ = $M_{0max}^{2}$ if $M_{0max}^{2}$ $\geq$0, and
$M_{0}^{2}$ = 0 if $M_{0max}^{2}$ $<$0.
We calculated $P_{0l}$:
\begin{equation}
   P_{0l} = \frac {\textstyle M_{B}^{2}-(M_{ch}^{2}+P_{t}^{2})-(M_{0}^{2}+P_{t}^{2})}{\textstyle 2 (M_{ch}^{2}+P_{t}^{2})} P_{chl},
\label{eqn:p0l} 
\end{equation}
and hence $E_{0}$ (Equation~(\ref{eqn:e0})).  
We then reconstructed the $B$ hadron energy, 
$E_{B}^{rec}=E_{0}+E_{ch}$.
Events were retained in which 
both reconstructed $B$ energies satisfied $E_B^{rec}<60$ GeV.
A final sample of 19,809 events was obtained with an estimated \bb selection
efficiency of 21.7\% and a background contribution of only 0.23\%, which
was almost entirely from \z0 \ra \cc events.

\begin{figure}[ht]      
\vskip 5truecm
\epsfysize5.0 in
\epsfxsize5.5 in
\epsfbox{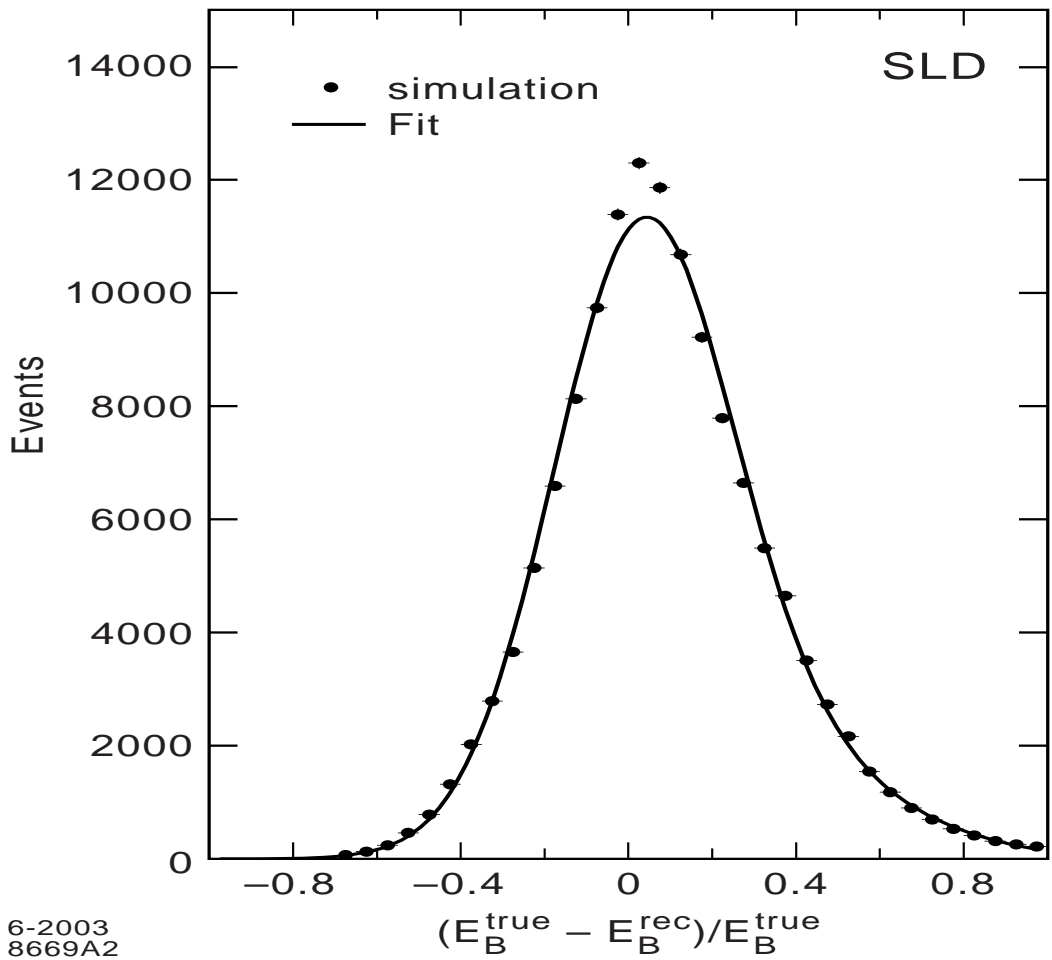}
\caption{Resolution on the reconstructed $B$-hadron energy.}
\label{fig:energyres}
\end{figure}

The energy resolution of the final $B$ sample is shown in 
Fig.~\ref{fig:energyres}, where we plot the normalised residual on $E_B$: 
$(E_B^{true}-
E_B^{rec})/E_B^{true}$.
The distribution was fitted to a double Gaussian function for which
the mean positions, widths and normalisations were allowed to vary. 
79.5\% of the
population lies in the `core' Gaussian, of width 21.3\%; the remaining
population is characterised by a `tail' Gaussian of width 31.3\%.

\begin{figure}[ht]      
\vskip 5truecm
\epsfysize5.0 in
\epsfxsize5.5 in
\epsfbox{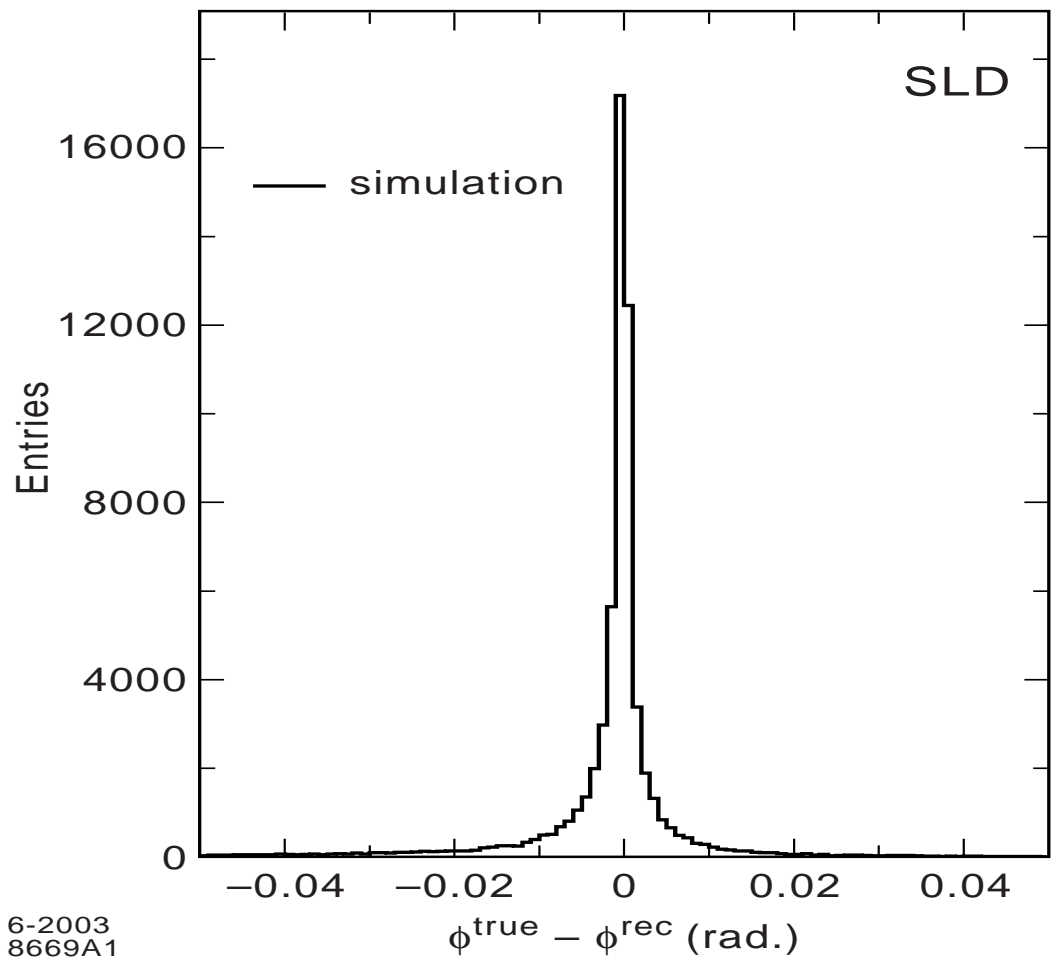}
\caption{Resolution on the reconstructed angle between the two $B$ hadrons.}
\label{fig:angleres}
\end{figure}

The angular resolution was similarly investigated and is shown in 
Fig.~\ref{fig:angleres}, where we plot the residual on $\phi$: $\phi^{true}-
\phi^{rec}$. The vast majority of angles are reconstructed to better than 
5 mrad. The tail in the resolution function corresponds to
$B$ decays with shorter decay lengths, where the vertex position error had a
larger relative effect on the determination of $\phi$.

\begin{figure}[ht]      
\vskip 5truecm
\epsfysize5.0 in
\epsfxsize5.5 in
\epsfbox{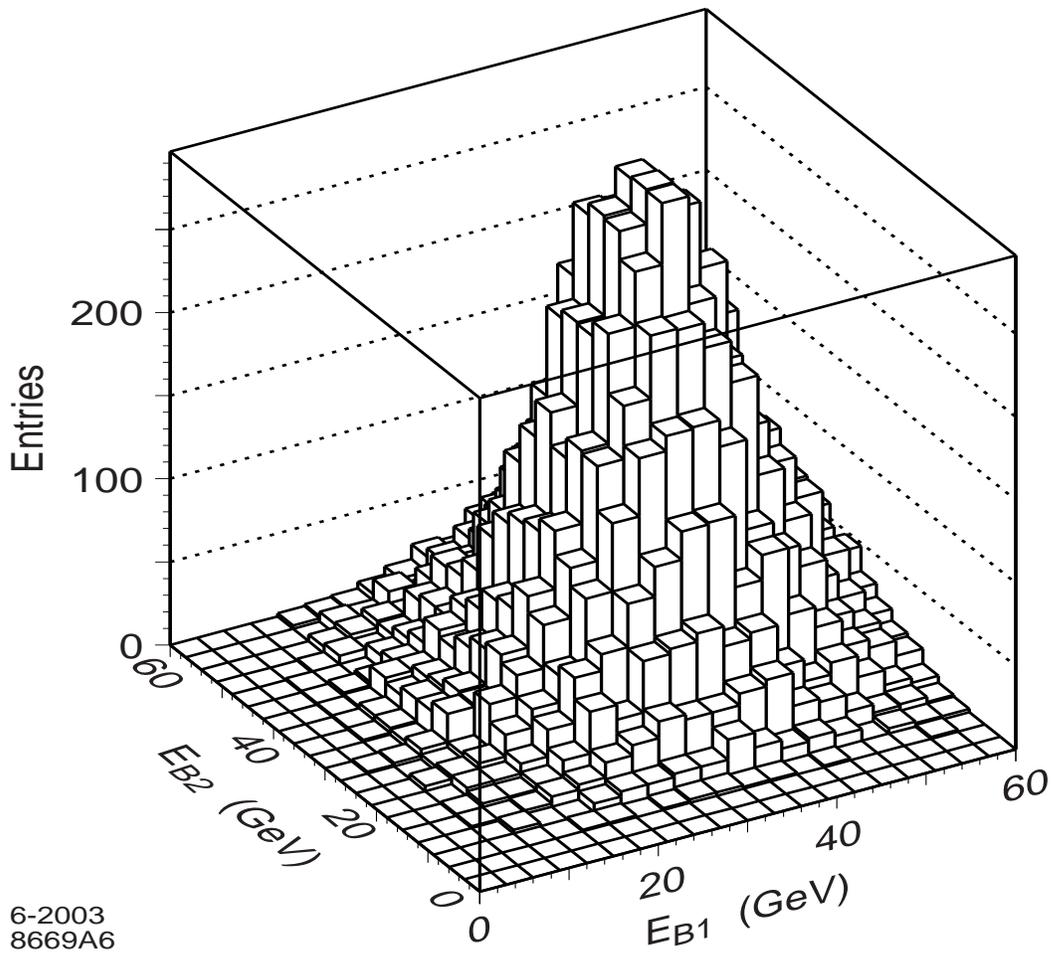}
\caption{Reconstructed double-inclusive $B$-hadron energy distribution.}
\label{fig:raw2d}
\end{figure}

The double-inclusive distribution of raw $B$-hadron energies is shown
in Figure~\ref{fig:raw2d}. Since we do not distinguish between $B$ and
$\overline{B}$ hadrons, the reconstructed $B$-decay vertices were labelled
arbitrarily `1' and `2' on an event-by-event basis.

We divided $E_B^{rec}$ by the beam energy, $E_{beam}=\sqrt{s}/2$, 
to obtain the reconstructed scaled $B$-hadron energy, 
$x_{B}^{rec}=E_{B}^{rec}/E_{beam}$.

\section{Angle-Dependent $B$-$\overline{B}$
Energy Moments}

\begin{figure}[ht]      
\vskip 5truecm
\epsfysize5.0 in
\epsfxsize5.5 in
\epsfbox{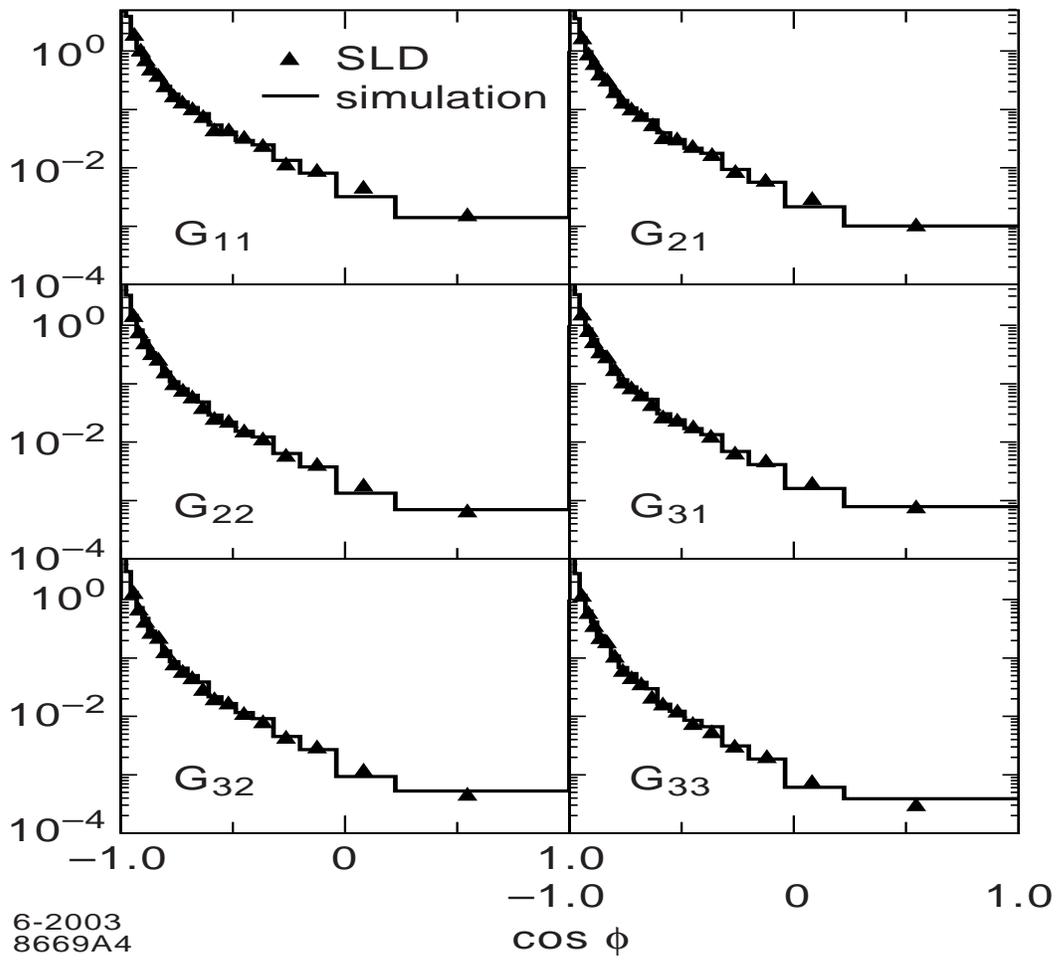}
\caption{Reconstructed moments of the two $B$-hadron energies 
(see text); the points include statistical error bars.}
\label{fig:gijraw}
\end{figure}

In each event
we quantified the correlations between the two $B$ hadrons in terms 
of the angle dependent scaled-energy moments proposed in \cite{phil}.
We formed the single-inclusive $B$-energy distribution 
and evaluated the moments (Eq.~\ref{eqn:single}) 
from the raw measured distribution:
\begin{equation}
D_i^{rec}\quad=\quad\frac{1}{2N}\Sigma_k (x_B^{rec})^{i-1}
\end{equation}
where the index $k$ ($1\leq k\leq 20$) runs over the bins of $x_B^{rec}$
and $N$ is the number of events in the sample. 
Similarly, from the double-inclusive $B$-energy distribution 
(Fig.~\ref{fig:raw2d}) we evaluated in each cos$\phi$ bin the double moments 
(Eq.~\ref{eqn:double}):
\begin{equation}
D_{ij}^{rec}(\phi)\quad=\quad\frac{1}{N}\Sigma_{N(\phi)} (x_{B1}^{rec})^{i-1}
(x_{B2}^{rec})^{j-1}
\end{equation}
where the sum extends over the set of events in each cos$\phi$ bin.
The normalised moments $G_{ij}$ (Eq.~\ref{eqn:norm}) were evaluated:
\begin{equation}
G_{ij}^{rec}(\phi)\quad=\quad\frac{D_{ij}^{rec}(\phi)}{D_i^{rec} D_j^{rec}}
\end{equation}
The first six moments, $i$ = 1,2,3 and $j$ = 1,$\ldots$,$i$ are shown in
Fig.~\ref{fig:gijraw}. The bin centers were defined by taking the 
average value of cos$\phi$ within each bin.

Also shown in Fig.~\ref{fig:gijraw} is a comparison with the simulated
normalised moments; the simulation reproduces the data.
Given that we showed previously~\cite{sldoldbfrag,sldnewbfrag} that the
Peterson function implemented in our simulation does not provide a good
description of the $b$-quark fragmentation function, this agreement may
naively appear to be surprising. However, if, as proposed in~\cite{phil}, 
the non-perturbative contributions to the normalised quantity $G_{ij}$ cancel, 
the agreement should be excellent, as observed.

We used our simulated event sample
 to correct for the effects of the detector acceptance, 
the efficiency of the technique for reconstructing $B$-hadron decays,  
the energy resolution, 
and bin migrations caused by the finite angular resolution.  We defined
a binwise correction factor:
\begin{equation}
F_{ij}^{MC}(\phi)\quad\equiv\quad \frac{G_{ij}^{gen}(\phi)|_{MC}}
{G_{ij}^{rec}(\phi)|_{MC}}
\end{equation}
where $G_{ij}^{gen}(\phi)|_{MC}$ is the normalised moment calculated using
generated true \ep \ra $B\overline{B}$ + $X$ events, and
 $G_{ij}^{rec}(\phi)|_{MC}$ is the corresponding moment calculated after
 simulation of the detector and application of the same analysis as applied to 
the data.
For cos$\phi$ $\sim$ $-1$ the value of $F_{ij}$ is close to unity. As 
cos$\phi$ increases towards 1, $F_{ij}$ rises monotonically to 
approximately 8 ($F_{11}$) or 2 ($F_{33}$). The increase with cos$\phi$
reflects our decreasing efficiency to select $B\overline{B}$ events as the
angle between the two $B$-decay vertices becomes smaller and the 
$B$ energies decrease. For a given 
cos$\phi$ bin, $F_{ij}$ decreases as $i$ and $j$ increase due to the 
effective weighting of this efficiency by $x_B^{i-1}x_B^{j-1}$.

We derived the true normalised moments:
\begin{equation}
G_{ij}(\phi)\quad=\quad F_{ij}^{MC}(\phi)\;G_{ij}^{rec}(\phi) 
\end{equation}
which are shown in Fig~\ref{fig:gij}.
The uncertainties associated with this correction procedure are discussed in 
the next section.

\section{Error Estimation}
\label{sec:sys}

We used our simulation to estimate the statistical error on the moments.
The simulated event sample was divided into subsamples, each comprising the
same number of events as in the data sample. The entire analysis was
performed on each of these subsamples. Within each cos$\phi$ bin the r.m.s. 
deviation of the ensemble of results was evaluated and taken as the 
statistical error
in that bin. For a given pair of $i$ and $j$ values the errors on the
particular moment are not correlated between different bins of cos$\phi$, 
but within
a given cos$\phi$ bin the errors on the different moments are correlated,
since the same set of events was used.

We considered sources of systematic uncertainty that potentially affect 
our measurements. 
These may be divided into uncertainties in 
modeling the detector and uncertainties on 
experimental measurements serving as
input parameters to the underlying physics modeling. 
For these studies our simulation was used.

Since our
energy reconstruction technique is strongly dependent on charged-track 
properties,
four sources of uncertainty were investigated:
our simulated tracking efficiency, transverse momentum resolution, and the
resolutions on the track impact parameter and polar angles.
In each case the simulation was corrected so as to 
reproduce the data. The full analysis was repeated on the data, and half 
the difference between the results obtained using the corrected and 
uncorrected simulations was taken as a symmetric systematic error.

A large number of measured quantities relating to the production and decay
of charm and bottom hadrons are used as input to our simulation. 
In \bb events we considered the uncertainties on: 
the branching fraction for \z0 \ra \bb;
the rates of production of $B^{\pm}$, $B^0$ and $B^0_s$ mesons, 
and $B$ baryons;
the lifetimes of $B$ mesons and baryons;
and the average $B$-hadron decay charged multiplicity.
In \cc events we considered the uncertainties on: 
the branching fraction for \z0 \ra \cc;
the charmed hadron lifetimes,
the charged multiplicity of charmed hadron decays,
the production of  $K^0$ from charmed hadron decays,
and the fraction of charmed hadron decays containing no $\pi^0$s.
We also considered the rates of production of 
secondary \bb and \cc from gluon splitting.
The uncertainty on the world-average value~\cite{sldnewbfrag} 
of each quantity was used to 
rederive the corrected $G_{ij}(\phi)$. In each bin of $G_{ij}(\phi)$ the
deviation between the rederived and standard values was taken as an
estimate of the corresponding systematic error.
Other relevant systematic effects such as variation of 
the event selection cuts and the assumed $B$-hadron mass were  
found to be very small.  

The error associated with the choice of $b$-quark fragmentation function used
in the simulation was estimated by rederiving the corrected data using in turn
the UCLA, Kartvelishvili and Bowler fragmentation 
functions~\cite{sldnewbfrag}. 
In each cos$\phi$ bin the r.m.s. deviation of the  results w.r.t. the 
standard
value, using the Peterson function, was taken as an estimate of the systematic 
error. This error was typically much smaller than that arising from the other
error sources.

For each systematic error investigated, the shape of the
angular dependence of the moments was not significantly 
changed, only the overall normalization was affected.  
For each moment, in each cos$\phi$ bin the dominant errors were typically 
those related to the uncertainties on the charged-track properties. The 
errors due to charm and bottom physics modeling were typically an order of 
magnitude smaller.  In each cos$\phi$ bin  all sources of systematic 
uncertainty were added in quadrature to obtain the total systematic error.
The fully-corrected $G_{ij}$, with statistical and systematic
errors added in quadrature, are shown in Fig.~\ref{fig:gij}.

\section{Comparison with Perturbative QCD Calculations}

\begin{figure}[ht]      
\vskip 5truecm
\epsfysize5.0 in
\epsfxsize5.5 in
\epsfbox{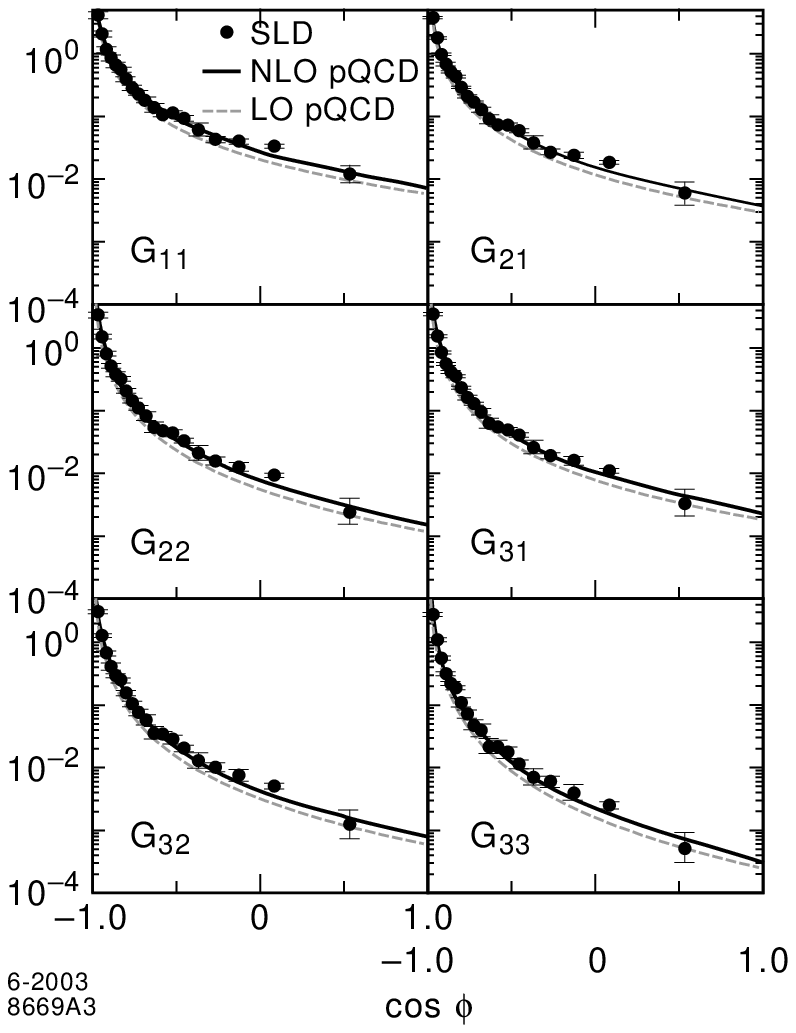}
\caption{Corrected moments of the two $B$-hadron energies compared
with LO and NLO pQCD calculations. The error bars are the sum in quadrature of 
the statistical and systematic errors (see text).}
\label{fig:gij}
\end{figure}

The fully-corrected data were compared (Fig.~\ref{fig:gij}) with a recent
calculation~\cite{bno} of the normalised moments complete at NLO in pQCD. 
Both the LO and full NLO calculations are shown in Fig.~\ref{fig:gij};
the calculations assume an \alpmzsq value of 0.120 and a pole $b$-quark mass
value of 5.0 GeV/$c^2$. 
It can be seen that the difference between the two is relatively small, and 
that the 
LO calculation lies systematically slightly below the NLO calculation. The 
small size of the NLO relative to the LO contributions is an indication 
that the normalised moments are perturbatively robust observables.  
For each moment shown, the LO calculation undershoots the data. The
NLO calculation reproduces the data across the full range of cos$\phi$.

This comparison does not rely on any convolution of the pQCD 
calculations with models of the non-perturbative
hadronisation process.
Hence the excellent agreement between the pQCD calculations and the data  
verifies the ansatz of 
factorization between the perturbative and non-perturbative phases that forms
the basis for the pQCD calculation of heavy-hadron properties.

\section*{Acknowledgements}
We thank the personnel of the SLAC accelerator department and the
technical
staffs of our collaborating institutions for their outstanding efforts
on our behalf. We thank A.~Brandenburg, P.~Nason and C.~Oleari for helpful
discussions.

\vskip .5truecm
\small
\vbox{\footnotesize\renewcommand{\baselinestretch}{1}\noindent
$^*$Work supported by Department of Energy
  contracts:

  DE-FG02-91ER40676 (BU),
  DE-FG03-91ER40618 (UCSB),
  DE-FG03-92ER40689 (UCSC),

  DE-FG03-93ER40788 (CSU),
  DE-FG02-91ER40672 (Colorado),
  DE-FG02-91ER40677 (Illinois),

  DE-AC03-76SF00098 (LBL),
  DE-FG02-92ER40715 (Massachusetts),
  DE-FC02-94ER40818 (MIT),

  DE-FG03-96ER40969 (Oregon),
  DE-AC03-76SF00515 (SLAC),
  DE-FG05-91ER40627 (Tennessee),

  DE-FG02-95ER40896 (Wisconsin),
  DE-FG02-92ER40704 (Yale);

  National Science Foundation grants:

  PHY-91-13428 (UCSC),
  PHY-89-21320 (Columbia),
  PHY-92-04239 (Cincinnati),

  PHY-95-10439 (Rutgers),
  PHY-88-19316 (Vanderbilt),
  PHY-92-03212 (Washington);

  The UK Particle Physics and Astronomy Research Council
  (Bristol, Brunel, QMUL, Oxford, RAL);

  The Istituto Nazionale di Fisica Nucleare of Italy

  (Bologna, Ferrara, Frascati, Pisa, Padova, Perugia);

  The Japan-US Cooperative Research Project on High Energy Physics
  (Nagoya, Tohoku);

  The Korea Research Foundation (Soongsil, 1997).}




\vfill
\eject

\section*{$^{**}$List of Authors}
%
%
%
\begin{center}
\def\iAOMORI{$^{(1)}$}
\def\iBRI{$^{(2)}$}
\def\iBRUN{$^{(3)}$}
\def\iBU{$^{(4)}$}
\def\iCOLO{$^{(5)}$}
\def\iCSU{$^{(6)}$}
\def\iFERR{$^{(7)}$}
\def\iFRAS{$^{(8)}$}
\def\iJHU{$^{(9)}$}
\def\iLBL{$^{(10)}$}
\def\iMASS{$^{(11)}$}
\def\iMISSI{$^{(12)}$}
\def\iMIT{$^{(13)}$}
\def\iMOSCOW{$^{(14)}$}
\def\iNAGO{$^{(15)}$}
\def\iOREG{$^{(16)}$}
\def\iOXF{$^{(17)}$}
\def\iPERU{$^{(18)}$}
\def\iQMUL{$^{(19)}$}
\def\iRAL{$^{(20)}$}
\def\iRUTG{$^{(21)}$}
\def\iSLAC{$^{(22)}$}
\def\iSOONG{$^{(23)}$}
\def\iTENN{$^{(24)}$}
\def\iTOHO{$^{(25)}$}
\def\iUCSB{$^{(26)}$}
\def\iUCSC{$^{(27)}$}
\def\iVAND{$^{(28)}$}
\def\iWASH{$^{(29)}$}
\def\iWISC{$^{(30)}$}
\def\iYALE{$^{(31)}$}

  \baselineskip=.75\baselineskip
\mbox{Koya Abe\unskip,\iTOHO}
\mbox{Kenji Abe\unskip,\iNAGO}
\mbox{T. Abe\unskip,\iSLAC}
\mbox{I. Adam\unskip,\iSLAC}
\mbox{H. Akimoto\unskip,\iSLAC}
\mbox{D. Aston\unskip,\iSLAC}
\mbox{K.G. Baird\unskip,\iMASS}
\mbox{C. Baltay\unskip,\iYALE}
\mbox{H.R. Band\unskip,\iWISC}
\mbox{T.L. Barklow\unskip,\iSLAC}
\mbox{J.M. Bauer\unskip,\iMISSI}
\mbox{G. Bellodi\unskip,\iOXF}
\mbox{R. Berger\unskip,\iSLAC}
\mbox{G. Blaylock\unskip,\iMASS}
\mbox{J.R. Bogart\unskip,\iSLAC}
\mbox{G.R. Bower\unskip,\iSLAC}
\mbox{J.E. Brau\unskip,\iOREG}
\mbox{M. Breidenbach\unskip,\iSLAC}
\mbox{W.M. Bugg\unskip,\iTENN}
\mbox{D. Burke\unskip,\iSLAC}
\mbox{T.H. Burnett\unskip,\iWASH}
\mbox{P.N. Burrows\unskip,\iQMUL}
\mbox{A. Calcaterra\unskip,\iFRAS}
\mbox{R. Cassell\unskip,\iSLAC}
\mbox{A. Chou\unskip,\iSLAC}
\mbox{H.O. Cohn\unskip,\iTENN}
\mbox{J.A. Coller\unskip,\iBU}
\mbox{M.R. Convery\unskip,\iSLAC}
\mbox{V. Cook\unskip,\iWASH}
\mbox{R.F. Cowan\unskip,\iMIT}
\mbox{G. Crawford\unskip,\iSLAC}
\mbox{C.J.S. Damerell\unskip,\iRAL}
\mbox{M. Daoudi\unskip,\iSLAC}
\mbox{N. de Groot\unskip,\iBRI}
\mbox{R. de Sangro\unskip,\iFRAS}
\mbox{D.N. Dong\unskip,\iMIT}
\mbox{M. Doser\unskip,\iSLAC}
\mbox{R. Dubois\unskip,}
\mbox{I. Erofeeva\unskip,\iMOSCOW}
\mbox{V. Eschenburg\unskip,\iMISSI}
\mbox{E. Etzion\unskip,\iWISC}
\mbox{S. Fahey\unskip,\iCOLO}
\mbox{D. Falciai\unskip,\iFRAS}
\mbox{J.P. Fernandez\unskip,\iUCSC}
\mbox{K. Flood\unskip,\iMASS}
\mbox{R. Frey\unskip,\iOREG}
\mbox{E.L. Hart\unskip,\iTENN}
\mbox{K. Hasuko\unskip,\iTOHO}
\mbox{S.S. Hertzbach\unskip,\iMASS}
\mbox{M.E. Huffer\unskip,\iSLAC}
\mbox{X. Huynh\unskip,\iSLAC}
\mbox{M. Iwasaki\unskip,\iOREG}
\mbox{D.J. Jackson\unskip,\iRAL}
\mbox{P. Jacques\unskip,\iRUTG}
\mbox{J.A. Jaros\unskip,\iSLAC}
\mbox{Z.Y. Jiang\unskip,\iSLAC}
\mbox{A.S. Johnson\unskip,\iSLAC}
\mbox{J.R. Johnson\unskip,\iWISC}
\mbox{R. Kajikawa\unskip,\iNAGO}
\mbox{M. Kalelkar\unskip,\iRUTG}
\mbox{H.J. Kang\unskip,\iRUTG}
\mbox{R.R. Kofler\unskip,\iMASS}
\mbox{R.S. Kroeger\unskip,\iMISSI}
\mbox{M. Langston\unskip,\iOREG}
\mbox{D.W.G. Leith\unskip,\iSLAC}
\mbox{V. Lia\unskip,\iMIT}
\mbox{C. Lin\unskip,\iMASS}
\mbox{G. Mancinelli\unskip,\iRUTG}
\mbox{S. Manly\unskip,\iYALE}
\mbox{G. Mantovani\unskip,\iPERU}
\mbox{T.W. Markiewicz\unskip,\iSLAC}
\mbox{T. Maruyama\unskip,\iSLAC}
\mbox{A.K. McKemey\unskip,\iBRUN}
\mbox{R. Messner\unskip,\iSLAC}
\mbox{K.C. Moffeit\unskip,\iSLAC}
\mbox{T.B. Moore\unskip,\iYALE}
\mbox{M. Morii\unskip,\iSLAC}
\mbox{D. Muller\unskip,\iSLAC}
\mbox{V. Murzin\unskip,\iMOSCOW}
\mbox{S. Narita\unskip,\iTOHO}
\mbox{U. Nauenberg\unskip,\iCOLO}
\mbox{H. Neal\unskip,\iYALE}
\mbox{G. Nesom\unskip,\iOXF}
\mbox{N. Oishi\unskip,\iNAGO}
\mbox{D. Onoprienko\unskip,\iTENN}
\mbox{L.S. Osborne\unskip,\iMIT}
\mbox{R.S. Panvini\unskip,\iVAND}
\mbox{C.H. Park\unskip,\iSOONG}
\mbox{I. Peruzzi\unskip,\iFRAS}
\mbox{M. Piccolo\unskip,\iFRAS}
\mbox{L. Piemontese\unskip,\iFERR}
\mbox{R.J. Plano\unskip,\iRUTG}
\mbox{R. Prepost\unskip,\iWISC}
\mbox{C.Y. Prescott\unskip,\iSLAC}
\mbox{B.N. Ratcliff\unskip,\iSLAC}
\mbox{J. Reidy\unskip,\iMISSI}
\mbox{P.L. Reinertsen\unskip,\iUCSC}
\mbox{L.S. Rochester\unskip,\iSLAC}
\mbox{P.C. Rowson\unskip,\iSLAC}
\mbox{J.J. Russell\unskip,\iSLAC}
\mbox{O.H. Saxton\unskip,\iSLAC}
\mbox{T. Schalk\unskip,\iUCSC}
\mbox{B.A. Schumm\unskip,\iUCSC}
\mbox{J. Schwiening\unskip,\iSLAC}
\mbox{V.V. Serbo\unskip,\iSLAC}
\mbox{G. Shapiro\unskip,\iLBL}
\mbox{N.B. Sinev\unskip,\iOREG}
\mbox{J.A. Snyder\unskip,\iYALE}
\mbox{H. Staengle\unskip,\iCSU}
\mbox{A. Stahl\unskip,\iSLAC}
\mbox{P. Stamer\unskip,\iRUTG}
\mbox{H. Steiner\unskip,\iLBL}
\mbox{D. Su\unskip,\iSLAC}
\mbox{F. Suekane\unskip,\iTOHO}
\mbox{A. Sugiyama\unskip,\iNAGO}
\mbox{A. Suzuki\unskip,\iNAGO}
\mbox{M. Swartz\unskip,\iJHU}
\mbox{F.E. Taylor\unskip,\iMIT}
\mbox{J. Thom\unskip,\iSLAC}
\mbox{E. Torrence\unskip,\iMIT}
\mbox{T. Usher\unskip,\iSLAC}
\mbox{J. Va'vra\unskip,\iSLAC}
\mbox{R. Verdier\unskip,\iMIT}
\mbox{D.L. Wagner\unskip,\iCOLO}
\mbox{A.P. Waite\unskip,\iSLAC}
\mbox{S. Walston\unskip,\iOREG}
\mbox{A.W. Weidemann\unskip,\iTENN}
\mbox{E.R. Weiss\unskip,\iWASH}
\mbox{J.S. Whitaker\unskip,\iBU}
\mbox{S.H. Williams\unskip,\iSLAC}
\mbox{S. Willocq\unskip,\iMASS}
\mbox{R.J. Wilson\unskip,\iCSU}
\mbox{W.J. Wisniewski\unskip,\iSLAC}
\mbox{J.L. Wittlin\unskip,\iMASS}
\mbox{M. Woods\unskip,\iSLAC}
\mbox{T.R. Wright\unskip,\iWISC}
\mbox{R.K. Yamamoto\unskip,\iMIT}
\mbox{J. Yashima\unskip,\iTOHO}
\mbox{S.J. Yellin\unskip,\iUCSB}
\mbox{C.C. Young\unskip,\iSLAC}
\mbox{H. Yuta\unskip.\iAOMORI}

\it
  \vskip \baselineskip                   
  \baselineskip=.75\baselineskip   
\iAOMORI
  Aomori University, Aomori, 030 Japan, \break
\iBRI
  University of Bristol, Bristol, United Kingdom, \break
\iBRUN
  Brunel University, Uxbridge, Middlesex, UB8 3PH United Kingdom, \break
\iBU
  Boston University, Boston, Massachusetts 02215, \break
\iCOLO
  University of Colorado, Boulder, Colorado 80309, \break
\iCSU
  Colorado State University, Ft. Collins, Colorado 80523, \break
\iFERR
  INFN Sezione di Ferrara and Universita di Ferrara, I-44100 Ferrara, Italy,
\break
\iFRAS
  INFN Laboratori Nazionali di Frascati, I-00044 Frascati, Italy, \break
\iJHU
  Johns Hopkins University,  Baltimore, Maryland 21218-2686, \break
\iLBL
  Lawrence Berkeley Laboratory, University of California, Berkeley, California
94720, \break
\iMASS
  University of Massachusetts, Amherst, Massachusetts 01003, \break
\iMISSI
  University of Mississippi, University, Mississippi 38677, \break
\iMIT
  Massachusetts Institute of Technology, Cambridge, Massachusetts 02139, \break
\iMOSCOW
  Institute of Nuclear Physics, Moscow State University, 119899 Moscow, Russia,
\break
\iNAGO
  Nagoya University, Chikusa-ku, Nagoya, 464 Japan, \break
\iOREG
  University of Oregon, Eugene, Oregon 97403, \break
\iOXF
  Oxford University, Oxford, OX1 3RH, United Kingdom, \break
\iPERU
  INFN Sezione di Perugia and Universita di Perugia, I-06100 Perugia, Italy,
\break
\iQMUL
  Queen Mary, University of London, London, E1 4NS, United Kingdom,
\break
\iRAL
  Rutherford Appleton Laboratory, Chilton, Didcot, Oxon OX11 0QX United Kingdom,
\break
\iRUTG
  Rutgers University, Piscataway, New Jersey 08855, \break
\iSLAC
  Stanford Linear Accelerator Center, Stanford University, Stanford, California
94309, \break
\iSOONG
  Soongsil University, Seoul, Korea 156-743, \break
\iTENN
  University of Tennessee, Knoxville, Tennessee 37996, \break
\iTOHO
  Tohoku University, Sendai, 980 Japan, \break
\iUCSB
  University of California at Santa Barbara, Santa Barbara, California 93106,
\break
\iUCSC
  University of California at Santa Cruz, Santa Cruz, California 95064, \break
\iVAND
  Vanderbilt University, Nashville,Tennessee 37235, \break
\iWASH
  University of Washington, Seattle, Washington 98105, \break
\iWISC
  University of Wisconsin, Madison,Wisconsin 53706, \break
\iYALE
  Yale University, New Haven, Connecticut 06511. \break

\rm
%

\end{center}

\vskip 1truecm
 
\vfill
\eject

\vfill
\eject

\vfill
\eject

\vfill
\eject

\vfill
\eject

\vfill
\eject

\end{document}